# Quantum noise limits in white-light-cavity-enhanced gravitational wave detectors

Minchuan Zhou,[1] Zifan Zhou,[2] Selim M. Shahriar,[1,2,*]

[1]Northwestern University, Department of Physics and Astronomy, 2145 Sheridan Road, Evanston, Illinois 60208, USA

[2]Northwestern University, Department of EECS, 2145 Sheridan Road, Evanston, Illinois 60208, USA

*shahriar@northwestern.edu

Previously, we had proposed a gravitational wave detector that incorporates the white-light-cavity (WLC) effect using a compound cavity for signal recycling (CC-SR). Here, we first use an idealized model for the negative dispersion medium (NDM) and use the so-called Caves model for a phase-insensitive linear amplifier to account for the quantum noise (QN) contributed by the NDM, in order to determine the upper bound of the enhancement in the sensitivity-bandwidth product. We calculate the quantum noise limited sensitivity curves for the CC-SR design, and find that the broadening of sensitivity predicted by the classical analysis is also present in these curves, but is somewhat reduced. Furthermore, we find that the curves always stay above the standard quantum limit. To circumvent this limitation, we modify the dispersion to compensate the non-linear phase variation produced by the optomechanical resonance effects. We find that the upper bound of the factor by which the sensitivity-bandwidth product is increased, compared to the highest-sensitivity result predicted by Bunanno and Chen [Phys. Rev. D 64, 042006 (2001)], is ~14. We also present a simpler scheme (WLC-SR) where a dispersion medium is inserted into the SR cavity. For this scheme, we found the upper bound of the enhancement factor to be ~18. We then consider an explicit system for realizing the NDM, which makes use of five energy levels in M-configuration to produce gain, accompanied by electromagnetically induced transparency (the GEIT system). For this explicit system, we employ the rigorous approach based on Master

Equation to compute the QN contributed by the NDM, thus enabling us to determine the enhancement in the sensitivity-bandwidth product definitively rather than the upper bound thereof. Specifically, we identify a set of parameters for which the sensitivity-bandwidth product is enhanced by a factor of 17.66.

## I. INTRODUCTION

Gravitational waves (GWs) modulate space-time, and for a specific polarization of GW, the modulations along two perpendicular axes are exactly out of phase [1]. Thus, the geometric configuration of a Michelson interferometer makes it a natural candidate for a GW detector. Under conditions where the excess noise is actively suppressed sufficiently, the performance of the detector is limited by quantum noise (QN), consisting of photon shot noise and radiation pressure noise [2]. For the first-generation Laser Interferometer Gravitational Wave Observatory (LIGO), where these two kinds of noises are uncorrelated, the Heisenberg uncertainty principle sets a standard quantum limit (SQL) for the minimum detectable gravitational-wave signal [3]. The Advanced LIGO (aLIGO) uses a combination of improved techniques. Along with power recycling (PR) and a higher power laser source, the aLIGO will employ signal recycling (SR) [4]. There are two special modes of operation corresponding to specific choices of reflectivity ($r_{SR}$) of the SR mirror ($M_{SR}$) and length of the SR cavity (SRC) formed ($L_{SRC}$) (Fig. 1) [5]: extreme signal recycling mode (narrowband operation) when $r_{SR}$ is high and $\varphi_{SRC} = k_c L_{SRC} (\mod 2\pi) = 0$ and extreme signal extraction mode (broadband operation) when $r_{SR}$ is low and $\varphi_{SRC} = \pi/2$ [4,6]. In both cases, the QN is above the SQL since the correlation between shot noise and radiation pressure noise is still zero. However, under modes when $\varphi_{SRC} \neq 0$ or $\pi/2$, a dynamical correlation between the two kinds of noises is created by $M_{SR}$. As a result, the QN can beat the SQL by roughly a factor of 2 over a small frequency range [7]. The dips in the noise curves correspond to optomechanical (OM) resonances [8]. The reflectivity $r_{SR}$ can be increased to create steeper dips with decreasing width. As we know, the GW signal is usually a chirp signal, with the frequency of interest 10Hz~$10^3$Hz. Thus the narrow frequency range of these dips may be too small for many types of sources.

A white light cavity (WLC) [9,10,11] is an optical cavity with a high buildup factor yet a broad response. Previously, we proposed a scheme (Fig. 8 in Ref. 9) for using the WLC effect to broaden the response of a GW detector without a reduction in sensitivity. In this design, which can be adapted to the aLIGO design relatively simply and noninvasively, we replace the conventional SR mirror with a compound cavity (CC) consisting of two mirrors and a negative dispersion medium (NDM). In what follows, we will refer to this as the CC-SR (compound cavity signal recycling) design, which will be reviewed briefly in Sec. II. When the dispersion is tuned to a critical value, the transmission window gets broadened significantly, without a reduction in the transmissivity. We have experimentally demonstrated a WLC in rubidium [11], and we have also explored a candidate system for producing this effect at the working wavelength for aLIGO [12].

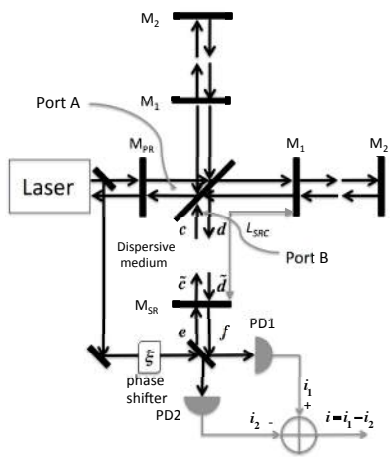

FIG. 1. Michelson interferometer with arm cavities and dual recycling (referred to as SR configuration) with homodyne detection.

The analysis presented in Ref. 9 for the CC-SR design did not take into account the effect of the QN. In this paper, we first augment this analysis in order to determine the QN-limited sensitivity of this architecture, using the two-photon formalism of Ref. 13, and an extension of the method employed in Refs. 7 and 14. We then modify the dispersion of the medium to compensate for the nonlinear phase variation induced by the OM effects. We also propose an alternative and simpler design (WLC-SR), where a dispersive medium with critically tuned dispersion is inserted in the SRC to achieve the phase compensation required by the OM resonance, and we analyze its QN-limited sensitivity. We present different cases where either a negative or positive dispersion medium (PDM) is used to cancel the phase variation, depending on the center frequency of the dispersion.

Since the QN predicted by the Caves model [15] is always less than or equal to that predicted by the ME model, initially we choose to consider an idealized NDM and make use of the Caves model in order to determine an upper bound of the degree of enhancement in the sensitivity-bandwidth product. The effect of the QN from the NDM is then included via a frequency-dependent gain/loss factor in the medium by taking into account an explicit model for the NDM. However, this model makes some assumptions that may not necessarily hold for some systems. To be more exact, we then use the Master Equation (ME) approach to calculate the QN from the NDM, which is established in Ref. 16, for a five-level, M-configuration Gain-EIT (GEIT) system for realizing the NDM.

The remainder of this paper is organized as follows: Section II describes in detail the CC-SR design, considering an explicit model for the dispersive medium, and analyzes the classical frequency response. Section III discusses the QN modeling with and without the excess QN from the dispersive medium taken into account, where the Caves model is used. Section IV discusses the effect of the NDM on the noise density curves in the CC-SR and also introduces a modification to the dispersion in order to achieve a broad region below the SQL. Section V introduces the WLC-SR configuration, exhibiting a broad sub-SQL region. In Sec. VI, we consider the GEIT system as the NDM and use the ME to calculate the QN from the GEIT system. We conclude in Sec. VII with a summary of our results and future plans. In the Appendix, we summarize the abbreviations used in the paper.

## II. CC-SR DESIGN AND ITS CLASSICAL RESPONSE

### A. CC-SR design

The Michelson interferometer with arm cavities and dual recycling is depicted in Fig. 1, referred to as the SR configuration. The interferometer is biased so that Port A is the bright port and Port B is the dark port [17]. Here the power recycling (PR) mirror ($M_{PR}$) in Port A and each of the front mirrors ($M_1$) form a PR cavity (PRC) with a length $L_{PRC}$ tuned so that it is a highly reflective compound mirror at the carrier frequency. The carrier light then resonates in the cavity established by each of the end mirrors ($M_2$) and the PRC. The net effect of $M_{PR}$ is to increase the effective power inside the two arms. In our analysis, we do not consider $M_{PR}$ explicitly [7,14]. Instead, we assume that a power much higher by a factor given by the finesse of the PRC than the laser output is entering the interferometer. $M_{SR}$ is inserted in Port B after the beam splitter and before the detector, forming an SRC with each of $M_1$. One of the sidebands produced by a monochromatic GW resonates in the cavity formed by $M_2$ and the SRC, thus producing an enhanced sensitivity around the resonant frequency with some bandwidth. In the homodyne detection scheme [18], the output from the interferometer is mixed with a local oscillator (LO) (which is produced by passing a piece of the carrier field through a phase shifter) at the beam splitter, and then detected with two photodetectors (PDs). The two resulting photocurrents $i_{1,2}$ are subtracted to obtain the final signal $i$.

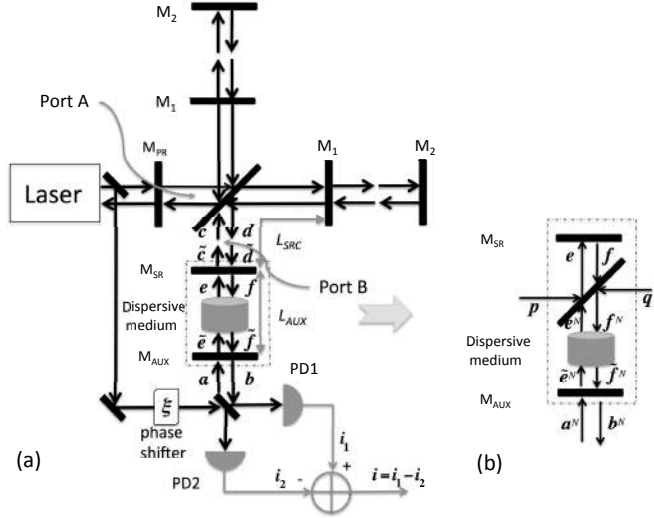

FIG. 2. (a) CC-SR design. In the dot-dashed box is the CC for SR, formed by the original SR mirror ($M_{SR}$) with a modified reflectivity matching that of $M_1$, $M_{AUX}$ and an NDM, replacing the single SR mirror. (b) Schematic view of the CC with excess QN from the NDM modeled by inserting a beam splitter with power reflectivity $R_{BS}$ and power transmissivity $T_{BS}$. Here $p$ and $q$ are the vacuum noises that leak into the system.

In the CC-SR design, summarized in Fig. 2, we modify the above configuration as follows. First, the reflectivity of $M_{SR}$ is changed to a value that matches the same for $M_1$. Next, we reduce the length of the SRC ($L_{SRC}$) by a factor of 20 (to ~0.5m), and tune it to be an integer multiple of the carrier wavelength. Under this condition, the transmissivity of the SRC becomes close to unity for a rather large range of frequencies around the carrier frequency. Thus, effectively, the SRC disappears for the range of GW sidebands we are interested in. Then we add an auxiliary mirror ($M_{AUX}$) for SR, and operate at the detuned mode where one of the sidebands resonates. To achieve a high degree of sensitivity, the reflectivity of $M_{AUX}$ is chosen to be fairly high, as a result of which the finesse of the cavity formed by $M_{AUX}$ and $M_2$ is very large, and the bandwidth of the sensitivity is narrowed. To compensate for this, we insert, between $M_{SR}$ and $M_{AUX}$, a medium with a critically tuned negative dispersion. Then $M_{AUX}$, $M_2$ and the NDM form a WLC. When the dispersion is tuned to the condition where the wavelength becomes independent of frequency over some bandwidth, the transmission profile of the cavity becomes much broader than that for the empty cavity. Thus, a broad bandwidth of the detector can be achieved while keeping the high degree of sensitivity.

The dispersion is designed such that the round-trip phase $\vartheta_{rt}$ gained by the light for any frequency is constant for a band around the resonant frequency $\omega_{res}$. For a dispersive medium with index of refraction $n(\omega)$ and length $l$ placed into a cavity of length $L$, the round-trip phase can be expressed in general as

$$\vartheta_{rt} = 2k(L-l) + 2n(\omega)kl + \vartheta_{ref}, \tag{1}$$

where $\vartheta_{ref}$ is the phase from reflection, and $k$ is the free space wave number. The WLC condition is satisfied if

$$\left.\frac{d\vartheta_{rt}}{d\omega}\right|_{\omega_{res}} = 0. \tag{2}$$

Assuming $n(\omega_{res}) = 1$, the condition above is equivalent to

$$\left.\frac{dn}{d\omega}\right|_{\omega_{res}} = -\frac{L}{l}\frac{1}{\omega_{res}}, \tag{3}$$

corresponding to a group index $n_g = 1 - L/l$. When the NDM fills up the whole cavity ($L=l$), this condition corresponds to a vanishing group index.

We first use an idealized model for the NDM. We will consider a more explicit model when we take into account the additional QN later. We assume that the NDM has a transmission profile which is given by a narrow band dip on top of a much broader gain [19]. The real and imaginary parts of the susceptibility $\chi \equiv \chi' + i\chi''$ are as follows:

$$\chi'' = -\frac{G_e \Gamma_e^2}{\vartheta_e} + \frac{G_i \Gamma_i^2}{\vartheta_i}, \tag{4}$$

$$\chi' = \frac{2G_e(\omega - \omega_c)\Gamma_e}{\vartheta_e} - \frac{2G_i(\omega - \omega_c)\Gamma_i}{\vartheta_i}, \tag{5}$$

where $\omega_c$ is the center frequency of the dispersion, and $\vartheta_k = 2\Omega_k^2 + \Gamma_k^2 + 4(\omega - \omega_c)^2$ ($k = e$ or $i$). Here "$e$" stands for the broad gain and "$i$" for the narrow dip. We use two parameters $\xi_k = \wp_k^2/(\hbar^2 \Gamma_k)$ to define the Rabi frequencies $\Omega_k^2 \equiv \Gamma_k E^2 \xi_k$ and the gain parameters $G_k = \hbar N_k \xi_k / \varepsilon_0$ [19]. The complex index of refraction is then $n_c = \sqrt{1+\chi} \approx 1 + \chi'/2 + i\chi''/2$. Thus, the total propagation phase in the cavity is $\varphi_{NDM} = (1 + \chi'/2) k l_{CAV} \pmod{2\pi}$, and the gain/loss factor is given by $g = \exp(-\chi'' k l_{CAV}/2)$, where $l_{CAV}$ is the length of the cavity. The gain factor is greater (less) than unity for $\chi'' > 0$ ($\chi'' < 0$). In the limit of vanishing Rabi frequencies ($\Omega_k \to 0$), the bandwidths of the profiles are given by $\Gamma_k$, which are chosen to be $\Gamma_e/2\pi = 0.8 \text{MHz}$ and $\Gamma_i = 10^3 \Gamma_e$, with the other parameters chosen to satisfy the WLC condition of Eq. (3).

## B. Classical frequency response

We have considered classically the frequency response of the GW detector in Ref. 9. The propagation of the light at the frequency $\omega_0$ under the influence of a GW at the frequency $\Omega$ induces sidebands at frequencies $\omega_0 \pm \Omega$ [9,14]. Using the complex representation of the electromagnetic field, the total field at Port B is $\tilde{E}_\pm = E_\pm e^{i(\omega_0 \pm \Omega)t}$ for the component at frequency $\omega_0 \pm \Omega$. At the beam splitter the output $\tilde{E}_{out} = \tilde{E}_+ + \tilde{E}_-$ mixes with a small amount of the carrier frequency light $\tilde{E}_L = E_L e^{i(\omega_0 t + \eta)}$ and the beat signal is [9]

$$\delta I = \tilde{E}_L \tilde{E}_+^* + \tilde{E}_L^* \tilde{E}_+ + \tilde{E}_L \tilde{E}_-^* + \tilde{E}_L^* \tilde{E}_-, \tag{6}$$

which can be written in the form

$$\delta I = P \cos[\Omega(t - L/c)] + Q \sin[\Omega(t - L/c)], \tag{7}$$

where

$$P = P_+ + P_-, Q = Q_+ + Q_-, \tag{8}$$

$$P_\pm = 2E_0 E_L B \frac{\zeta_\pm \left[ r_2 r_{S\pm} \cos(2k_\pm L - \phi_{rS\pm} + \phi_{tS\pm} + \phi_C) - \cos(\phi_{tS\pm} + \phi_C) \right]}{1 + F'_{S\pm} \sin^2(k_\pm L_S - \phi_{rS\pm}/2)}, \tag{9}$$

$$Q_{\pm} = \mp 2 E_0 E_L B \frac{\zeta_{\pm}\left[-r_2 r_{S\pm} \sin(2 k_{\pm} L - \phi_{rS\pm} + \phi_{tS\pm} + \phi_C) + \sin(\phi_{tS\pm} + \phi_C)\right]}{1 + F'_{S\pm} \sin^2(k_{\pm} L_S - \phi_{rS\pm}/2)}. \tag{10}$$

The magnitude of the signal is given by $|\delta I| = \sqrt{P^2 + Q^2}$. Here $r_{S\pm} e^{i\phi_{rS\pm}}$ and $t_{S\pm} e^{i\phi_{tS\pm}}$ are the reflectivity and transmissivity, respectively, of the compound mirror $M_S$ composed of $M_1$, $M_{SR}$ and $M_{AUX}$ at the frequency $\omega_0 \pm \Omega$, and $k_{\pm} = (\omega_0 \pm \Omega)/c$ is the wave number. The two arm cavities are identical and of length $L$. $L_S$ is the distance from $M_S$ to $M_2$, which in the case of the CC-SR is the same as $L$. The relevant parameters are defined as

$$F'_{S\pm} = \frac{4 r_{S\pm} r_2}{(1 - r_{S\pm} r_2)^2}, \quad F'_C = \frac{4 r_1 r_2}{(1 - r_1 r_2)^2}, \tag{11}$$

$$\zeta_{\pm} = \frac{t_1 t_{S\pm} r_2 h \omega_0 \sin(\Omega L/c)}{\Omega(1 - r_1 r_2)^2 (1 - r_2 r_{S\pm})^2}, \quad B e^{i\phi_B} = \frac{e^{-2 i k_c L} - r_1 r_2}{1 + F'_C \sin^2(k_c L)}, \tag{12}$$

$$\phi_C = -2 k_c L - \eta + \phi_B - \pi/2. \tag{13}$$

When the NDM is inserted, $k_{\pm} L_S$ needs to be changed to $n(\omega_0 \pm \Omega) k_{\pm} L_S$ in the equations above.

In Fig. 3, we illustrate the effect of the WLC on the frequency response of the GW detector, which shows $|\delta I|$ as a function of $\Omega$ with $R_{AUX} = 99\%$, with the values of the other parameters being as shown in Table I. We assumed that $M_2$ is totally reflective ($r_2 = 1$ and $t_2 = 0$), while $M_1$ has reflectivity $r_1$ and transmissivity $t_1$. The reflectivity and transmissivity, respectively, are $r_{SR}$ and $t_{SR}$ for $M_{SR}$, and $r_{AUX}$ and $t_{AUX}$ for $M_{AUX}$. Without the NDM, when the length $L_{AUX}$ is chosen so that the sideband at $\omega_{res} = \omega_0 + \gamma$ is resonant in the cavity formed by $M_{AUX}$ and $M_2$, the response curve is peaked at $\Omega = \gamma$, with $P_+$ and $Q_+$ contributing most to $|\delta I|$, and $P_-$ and $Q_-$ being negligible, since only one of the sidebands (in this case it is the positive sideband at $\omega_0 + \gamma$) is on resonance. Adding the NDM, which has its dispersion centered at $\omega_{res}$ and the shape tailored according to Eq. (3), broadens the response curve with the amplitude being twice as large as that in the case without the NDM. However, the discussion above did not take into account the OM effects that modify the resonance condition, and, as will be shown in Sec. IV, the QN-limited sensitivity curves show significant broadening but remain above the SQL.

We note here that the results shown in Fig. 3 do not take into account enhancement of the sidebands due to the (spectrally varying) gain from the NDM. If this were to be considered, the enhancement in sensitivity would be even larger. However, as is well known now, the sensitivity profile obtained through such a semi-classical analysis is essentially irrelevant. What matters instead is the minimum detectable GW strain amplitude when the effects of QN and the OM resonance are taken into account. Thus, the results we derive later in this paper by taking the effects of QN (including those due to the gain of the NDM) and the OM effects are the more relevant ones.

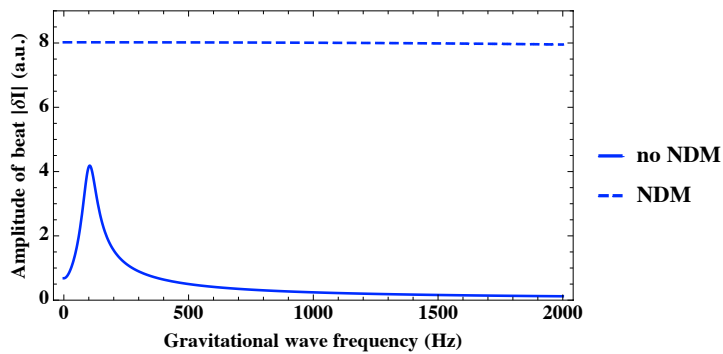

FIG. 3.    Response functions for the CC-SR design with $R_{AUX} = 99\%$, both with (plotted in red) and without (plotted in blue) the NDM inserted.

TABLE I. Values used in plotting Fig. 3.

| $k_c = 2\pi/(1064 \times 10^{-9})\text{m}^{-1}$ | $L = 2\pi(3.75446 \times 10^9)/k_c \approx 4\text{km}$ |
|---|---|
| $\omega_0 = k_c c \approx 1.77 \times 10^{15}\text{Hz}$ | $L_{SRC} = 2\pi(0.47 \times 10^6)/k_c \approx 0.50\text{m}$ |
| $t_{SR} = t_1 \approx 0.183, t_2 = 0$ | $k_{res} = k_c + \gamma/c$ |
| $t_{AUX} = 0.1$ | $L_{AUX} = -L + 2\pi(7.51 \times 10^9)/(2k_{res}) \approx 0.57\text{m}$ |

### III.   NOISE MODELING

Following the two-photon formalism developed by Caves and Schumaker [13], Kimble *et al.* have derived the input-output relations for a Michelson interferometer [14], and Buonanno and Chen have derived the input-output relation for an SR interferometer [7]. With the same formalism, we here develop the input-output relation for the CC-SR scheme. For simplicity, we consider first only the dispersive property of the NDM, and do not take into account the QN resulting from the gain spectrum. This is then followed by a more complete analysis where both the effects of the dispersion and the gain due to the NDM on the QN are taken into account.

## A. Input-output relation

We make the following assumptions. First, the length of the arm cavities $L$ oscillates at a frequency ~1Hz [20], around an equilibrium position where the laser frequency resonates (this effect in the final signal can be filtered by a high-frequency pass, thus the mirrors can be considered as still effectively). Second, the mirrors and beam splitters are lossless and infinitely thin. We describe the light with a time-dependent electric field, at fixed locations along the optical axis [14]. The laser as a carrier field enters the interferometer at the bright port with intensity $I_0$ and frequency $\omega_0$. The GW with a frequency $\Omega$ will interact with the carrier field to create sidebands at $\omega_0 \pm \Omega$ [9,14]. We denote the usual annihilation and creation operators for a photon at frequency $\omega$ by $a(\omega)$ and $a^\dagger(\omega)$. The amplitudes of the two-photon modes are defined as [13]

$$a_1(\Omega) = a(\omega_0 + \Omega)\sqrt{\frac{\omega_0 + \Omega}{2\omega_0}} + a^\dagger(\omega_0 - \Omega)\sqrt{\frac{\omega_0 - \Omega}{2\omega_0}}, \tag{14a}$$

$$a_2(\Omega) = -ia(\omega_0 + \Omega)\sqrt{\frac{\omega_0 + \Omega}{2\omega_0}} + ia^\dagger(\omega_0 - \Omega)\sqrt{\frac{\omega_0 - \Omega}{2\omega_0}}. \tag{14b}$$

$a_j(\Omega)\,(j=1,2)$ operates on two photons at frequencies $\omega_0 \pm \Omega$ simultaneously. If we ignore the terms proportional to $\Omega/\omega_0\,(\Omega/\omega_0 \ll 1)$, the commutation relations are [13]

$$\begin{aligned}[a_1(\Omega), a_2^\dagger(\Omega')] &= -[a_2(\Omega), a_1^\dagger(\Omega')] = 2\pi i \delta(\Omega - \Omega'), \\ [a_j(\Omega), a_j(\Omega')] &= 0 = [a_j(\Omega), a_j^\dagger(\Omega')], j=1,2.\end{aligned} \tag{15}$$

The electric field $E_a$ can then be expressed as a linear combination of two quadratures $E_{a1}$ and $E_{a2}$:

$$E_a(t) = E_{a1}(t)\cos(\omega_0 t) + E_{a2}(t)\sin(\omega_0 t), \tag{16}$$

$$E_{aj}(t) = \sqrt{\frac{4\pi\hbar\omega_0}{\mathcal{A}c}}\int_0^{+\infty}\left[a_j(\Omega)e^{-i\Omega t} + a_j^\dagger(\Omega)e^{i\Omega t}\right]\frac{d\Omega}{2\pi}, j=1,2, \tag{17}$$

where $\mathcal{A}$ is the effective cross-section area of the laser. For convenience, we use a vector $\boldsymbol{a}(\Omega) = \left(a_1(\Omega), a_2(\Omega)\right)^T$ to represent $E_a$ in the latter context. In order to establish the basic notation, we start by considering the simpler case, corresponding to the original LIGO, where there is no SR mirror, and the input field at Port A has a classical carrier field assumed to be only in the first quadrature. At Port B, the quadrature amplitudes for the input field are $c(\Omega)$, and those for the output are $d(\Omega)$. Following the method in Ref. 14, we derive the relations between $c(\Omega)$ and $d(\Omega)$:

$$d(\Omega) = A(\Omega)c(\Omega) + B(\Omega)\tilde{h}(\Omega), \tag{18}$$

where $A(\Omega)$ is a $2\times 2$ matrix and $B(\Omega)$ is a two-dimensional column vector with elements:

$$A_{11} = A_{22} = \frac{e^{2i\Omega L/c} - r_1}{1 - r_1 e^{2i\Omega L/c}}, A_{12} = 0, A_{21} = -\frac{I_0}{I_{SQL}}\frac{8L^2\gamma^4}{\Omega^2 c^2}\frac{e^{i\Omega L/c}}{(1 - r_1 e^{2i\Omega L/c})^2}, \tag{19a}$$

$$B_1 = 0, B_2 = \sqrt{\frac{I_0}{I_{SQL}}}\frac{4L\gamma^2}{\Omega c}\frac{e^{i\Omega L/c}}{1 - r_1 e^{2i\Omega L/c}}, \tilde{h}(\Omega) = \frac{h(\Omega)}{h_{SQL}(\Omega)}. \tag{19b}$$

Here $\gamma = t_1^2 c/(4L) = 2\pi \times 100$Hz is the half-bandwidth of the arm cavities [14]; $m = 40$kg is the mass of the mirrors; $I_{SQL} = mL^2\gamma^4/(4\omega_0) = 1.4\times 10^4$ W is the input light power for which the shot noise equals the radiation pressure noise at $\Omega = \gamma$; and $h_{SQL}(\Omega)$ is the SQL for $h(\Omega)$, which is the Fourier transform of the dimensionless GW signal $h(t) = \Delta L(t)/L$ [14]:

$$h_{SQL}(\Omega) = \sqrt{\frac{8\hbar}{m\Omega^2 L^2}}. \tag{20}$$

Under the assumptions that $\Omega L/c \ll 1$ and $t_1 \ll 1$, we have $e^{i\Omega L/c} \approx 1 + i\Omega L/c$ and $r_1 \approx 1 - t_1^2/2$. This yields $A_{11} = A_{22} \approx e^{2i\beta(\Omega)}$ and $A_{21} \approx -\mathcal{K}(\Omega)e^{2i\beta(\Omega)}$, where $\beta(\Omega) = \arctan(\Omega/\gamma)$ and $\mathcal{K}(\Omega) = 2I_0\gamma^4/\left[I_{SQL}\Omega^2(\Omega^2 + \gamma^2)\right]$, same as the results in Ref. 14.

Free-space propagation of the field operator is represented by a rotation of the vector $a(\Omega)$ with a phase shift [7]. For instance, if $\tilde{a}(\Omega)$ is the field after propagating a distance of $L$, then $\tilde{a}(\Omega) = \mathcal{R}(\varphi, \Phi)a(\Omega)$, where

$$\mathcal{R}(\varphi, \Phi) = e^{i\Phi}\begin{pmatrix} \cos\varphi & -\sin\varphi \\ \sin\varphi & \cos\varphi \end{pmatrix}, \tag{21}$$

with the phases $\varphi = \omega_0 L/c \pmod{2\pi}$ and $\Phi = \Omega L/c \pmod{2\pi}$. If the light travels through a dispersive medium with index of refraction $n(\omega)$, the matrix becomes

$$\mathcal{R}_n(\varphi,\Phi) = \frac{1}{2}\begin{pmatrix} e^{in(\omega_0+\Omega)(\varphi+\Phi)} + e^{-in(\omega_0-\Omega)(\varphi-\Phi)} & i\left[e^{in(\omega_0+\Omega)(\varphi+\Phi)} - e^{-in(\omega_0-\Omega)(\varphi-\Phi)}\right] \\ -i\left[e^{in(\omega_0+\Omega)(\varphi+\Phi)} - e^{-in(\omega_0-\Omega)(\varphi-\Phi)}\right] & e^{in(\omega_0+\Omega)(\varphi+\Phi)} + e^{-in(\omega_0-\Omega)(\varphi-\Phi)} \end{pmatrix}. \quad (22)$$

Since only one of the sidebands is at resonance and contributes to the output and the contribution of the other sideband is negligible as argued in Sec. II B, we make the approximation in the above equation that $n(\omega_0+\Omega) \approx n(\omega_0-\Omega)$ when designing the dispersion [we will also show the results using the exact matrix in Eq. (22) in the analysis later], and thus Eq. (22) becomes

$$\mathcal{R}_n(\varphi,\Phi) = e^{in(\omega_0+\Omega)\Phi}\begin{pmatrix} \cos[n(\omega_0+\Omega)\varphi] & -\sin[n(\omega_0+\Omega)\varphi] \\ \sin[n(\omega_0+\Omega)\varphi] & \cos[n(\omega_0+\Omega)\varphi] \end{pmatrix}. \quad (23)$$

Now we add a CC for SR, which is composed of $M_{SR}$, $M_{AUX}$ and an NDM, and assume that the distance from the beam splitter to $M_1$ is negligible compared to $L_{SRC}$. As shown in Fig. 2, $\tilde{c}(\Omega)$ and $\tilde{d}(\Omega)$ denote the fields just before $M_{SR}$, while $e(\Omega)$ and $f(\Omega)$ denote the fields immediately after $M_{SR}$; $\tilde{e}(\Omega)$ and $\tilde{f}(\Omega)$ denote the fields before $M_{AUX}$, while $a(\Omega)$ and $b(\Omega)$ denote the fields immediately after $M_{AUX}$. We define the rotation operators for propagation through $L_{SRC}$ and $L_{AUX}$ as $\mathcal{R}(\varphi_{SRC},\Phi_{SRC}) = \mathcal{R}_{SRC}$ and $\mathcal{R}_n(\varphi_{AUX},\Phi_{AUX}) = \mathcal{R}_{AUX}$, with $\varphi_{SRC} = \omega_0 L_{SRC}/c \,(\mathrm{mod}\, 2\pi)$, $\Phi_{SRC} = \Omega L_{SRC}/c \,(\mathrm{mod}\, 2\pi)$, $\varphi_{AUX} = \omega_0 L_{AUX}/c \,(\mathrm{mod}\, 2\pi)$, and $\Phi_{AUX} = \Omega L_{AUX}/c \,(\mathrm{mod}\, 2\pi)$. Here we assume that the NDM fills up the whole CC. The effect of the spectrally varying gain profile of the NDM will add additional QN. We will take this into account later on. For now, we ignore the effect of this gain [Fig. 2(a)]. In that case, we have the following relation:

$$c(\Omega) = \mathcal{R}_{SRC}\tilde{c}(\Omega), \tilde{d}(\Omega) = \mathcal{R}_{SRC}d(\Omega), e(\Omega) = \mathcal{R}_{AUX}\tilde{e}(\Omega), \tilde{f}(\Omega) = \mathcal{R}_{AUX}f(\Omega), \quad (24a)$$

$$\tilde{c}(\Omega) = t_{SR}e(\Omega) + r_{SR}\tilde{d}(\Omega), f(\Omega) = -r_{SR}e(\Omega) + t_{SR}\tilde{d}(\Omega), \quad (24b)$$

$$\tilde{e}(\Omega) = t_{AUX}a(\Omega) + r_{AUX}\tilde{f}(\Omega), b(\Omega) = -r_{AUX}a(\Omega) + t_{AUX}\tilde{f}(\Omega). \quad (24c)$$

Solving the system of Eq. (18) and Eqs. (24a)–(24c) gives the following input-output relation:

$$b(\Omega) = X(\Omega)a(\Omega) + Y(\Omega)\tilde{h}(\Omega), \quad (25)$$

where

$$U(\Omega) = t_{SR}^2 \mathcal{R}_{SRC} A(\Omega)[\mathcal{R}_{SRC}^{-1} - r_{SR}\mathcal{R}_{SRC} A(\Omega)]^{-1} - r_{SR}I, \tag{26a}$$

$$V(\Omega) = t_{SR}r_{SR}\mathcal{R}_{SRC} A(\Omega)[\mathcal{R}_{SRC}^{-1} - r_{SR}\mathcal{R}_{SRC} A(\Omega)]^{-1}\mathcal{R}_{SRC} B(\Omega) + t_{SR}\mathcal{R}_{SRC} B(\Omega), \tag{26b}$$

$$X(\Omega) = t_{AUX}^2 \mathcal{R}_{AUX} U(\Omega)[\mathcal{R}_{AUX}^{-1} - r_{AUX}\mathcal{R}_{AUX} U(\Omega)]^{-1} - r_{AUX}I, \tag{26c}$$

$$Y(\Omega) = t_{AUX}r_{AUX}\mathcal{R}_{AUX} U(\Omega)[\mathcal{R}_{AUX}^{-1} - r_{AUX}\mathcal{R}_{AUX} U(\Omega)]^{-1}\mathcal{R}_{AUX} V(\Omega) + t_{AUX}\mathcal{R}_{AUX} V(\Omega). \tag{26d}$$

Here $I$ is a 2×2 identity matrix, and $A(\Omega)$ and $B(\Omega)$ are as defined in Eqs. (19a) and (19b). It is confirmed that $X_{ij}$ has a common phase factor and $|\text{Det}(X)| = 1$, so $b_j(\Omega)$ ($j=1,2$) follows the same commutation relations as $a_j(\Omega)$ in Eq. (13). The relation for the SR configuration can be recovered by setting $r_{AUX} = 0$ and $L_{AUX} = 0$. In that case, we get $f(\Omega) = b(\Omega)$ and $e(\Omega) = c(\Omega)$. It should be noted that unlike in LIGO, where the GW signal $h(\Omega)$ only appears in the second quadrature of the output, $h(\Omega)$ now appears in both quadratures for SR (due to the presence of $M_{SR}$) and for CC-SR (due to the presence of $M_{SR}$ and $M_{AUX}$).

### B. Noise spectral density

At the beam splitter, the output $E_b(t) = E_{b1}(t)\cos(\omega_0 t) + E_{b2}(t)\sin(\omega_0 t)$ mixes with $E(t) = E_L \cos(\omega_0 t - \xi)$. The resulting photocurrents are then

$$i_1(t) \propto \overline{|E(t) + E_b(t)|^2} = \frac{1}{2}E_L^2 + \frac{1}{2}E_{b1}^2 + \frac{1}{2}E_{b2}^2 + E_L E_{b1}(t)\cos(\xi) + E_L E_{b2}(t)\sin(\xi), \tag{27a}$$

$$i_2(t) \propto \overline{|E(t) - E_b(t)|^2} = \frac{1}{2}E_L^2 + \frac{1}{2}E_{b1}^2 + \frac{1}{2}E_{b2}^2 - E_L E_{b1}(t)\cos(\xi) - E_L E_{b2}(t)\sin(\xi), \tag{27b}$$

where the horizontal bar indicates an averaging over a period much longer than $\omega_0^{-1}$. We detect the difference of $i_{1,2}$:

$$i(t) = i_1(t) - i_2(t) = 2E_L E_{b\xi}, \tag{28}$$

where we have defined

$$E_{b\xi}(\Omega) = E_{b1}(\Omega)\cos(\xi) + E_{b2}(\Omega)\sin(\xi), \tag{29}$$

which can also be expressed as

$$E_{b_\xi}(\Omega) = \sqrt{\frac{4\pi\hbar\omega_0}{Ac}} \int_0^{+\infty} \left[ b_\xi(\Omega)e^{-i\Omega t} + b_\xi^\dagger(\Omega)e^{i\Omega t} \right] \frac{d\Omega}{2\pi}, \tag{30}$$

where

$$b_\xi(\Omega) = b_1(\Omega)\cos(\xi) + b_2(\Omega)\sin(\xi). \tag{31}$$

The output $b_\xi(\Omega)$ consists of the signal component $\langle b_\xi(\Omega) \rangle$ and the noise component $\Delta b_\xi(\Omega)$:

$$\langle b_\xi(\Omega) \rangle = \left[ Y_1(\Omega)\cos(\xi) + Y_2(\Omega)\sin(\xi) \right] \frac{h}{h_{SQL}}, \tag{32}$$

$$\Delta b_\xi(\Omega) = \left[ X_{11}\sin(\xi) + X_{21}\cos(\xi) \right] a_1 + \left[ X_{12}\sin(\xi) + X_{22}\cos(\xi) \right] a_2. \tag{33}$$

The noise in the gravitational-wave signal $h$ at frequency $\Omega$ is related to the noise in the output $b_\xi(\Omega)$ via a transfer function

$$\Delta h(\Omega) = \frac{h_{SQL}(\Omega)}{Y_1(\Omega)\cos(\xi) + Y_2(\Omega)\sin(\xi)} \Delta b_\xi(\Omega). \tag{34}$$

Using the definition of spectral density [13]

$$2\pi\delta(\Omega - \Omega')S_h(\Omega) = \langle in | \Delta h(\Omega)\Delta h^\dagger(\Omega') + \Delta h^\dagger(\Omega')\Delta h(\Omega) | in \rangle, \tag{35}$$

and the fact that the input of the detector at the dark port is in the vacuum state $|in\rangle = |0_a\rangle$, we derive the noise spectral density for the GW signal $h(\Omega)$:

$$S_h^\xi(\Omega) \equiv h_n^2(\Omega) = h_{SQL}^2(\Omega) \frac{\left| X_{11}\sin(\xi) + X_{21}\cos(\xi) \right|^2 + \left| X_{12}\sin(\xi) + X_{22}\cos(\xi) \right|^2}{\left| Y_1 \sin(\xi) + Y_2 \cos(\xi) \right|^2}. \tag{36}$$

C. Inclusion of the QN from the NDM

In addition to the QN we have considered above, we must take into account the excess QN resulting from the NDM used for WLC. Specifically, this QN results from the fact that the NDM, in addition to providing dispersion, also amplifies or attenuates the signal. Physically, the noise associated in the amplification or attenuation is due to the spontaneous emission of photons that must accompany such a process. In deriving the QN from the NDM, we note first that the NDM is a phase-insensitive linear amplifier. For such an amplifier, it is in general possible to evaluate the QN by using the approach developed by Caves [15]. However, this model may not necessarily apply to complex systems. We have recently developed a rigorous approach based on Master Equation (ME) to determine QN in arbitrary complex atomic systems [16]. We have applied this approach to several different atomic systems, and compared the results to those predicted by the Caves model. We found that in most cases the Caves model is inadequate. However, we also found several examples where the prediction of the Caves model agrees with that of the ME approach. Furthermore, we found that the QN predicted by the Caves model is always less than or equal to what is predicted by the ME approach. In general, use of the ME approach is tedious and cumbersome. Thus, in this paper, we do not use the ME approach until Sec. VI, which contains the final findings of this paper, based on an explicit system for realizing the NDM. In the other sections, we make use of the simple Caves model in order to determine the upper bound of the enhancement in the sensitivity-bandwidth product achievable under various combinations of configurations and dispersion profiles. We do note, however, that in the result presented in Sec. V A, where we consider a positive dispersion medium (PDM) as the phase compensator, the prediction based on using the Caves model would be the same as that made using the ME model. This is because the PDM is generated by using an electromagnetically induced transparency (EIT) process employing a Λ-type three-level system, for which the Caves model agrees exactly with the ME approach, as shown in Ref. 16.

The Caves model can be described as follows. We define a factor $g$ as the intensity gain or loss factor of the dispersive medium:

$$g = \exp\left[-\chi''(\omega_0 + \Omega)kL_{AUX}\right]. \tag{37}$$

Generally, to account for the noise from a phase insensitive linear amplifier, a vacuum field $v(\Omega)$ is added after propagating through the medium:

$$y^*(\Omega) = \sqrt{g}\,y(\Omega) + \sqrt{g-1}\,v^\dagger(\Omega), g > 1, \tag{38}$$

while for the case of an attenuator,

$$y^*(\Omega) = \sqrt{g}\,y(\Omega) + \sqrt{1-g}\,v(\Omega), g < 1, \tag{39}$$

where $y$ ($y^*$) is the field operator before (after) propagating through the medium. Here $v^\dagger$ and $v$ are used so that the commutation relations for $y$ are maintained for $y^*$. Note also that in keeping with the approximation $n(\omega_0 + \Omega) \approx n(\omega_0 - \Omega)$ in Eq. (23), we approximate here $g(\omega_0 + \Omega) \approx g(\omega_0 - \Omega)$, which we will call later the single-sideband approximation (SSA). The exact results without the SSA are shown later.

For a general model that works both for gain and loss, we model the QN by placing inside the WLC a beam splitter (BS) that has a power reflectivity of $R_{BS}$ and power transmissivity of $T_{BS}$, from which the vacuum fields can leak into the system from the outside. We define

$$T_{BS}(\Omega) = g, R_{BS}(\Omega) = |1 - T_{BS}(\Omega)|, \tag{40}$$

and we write

$$y^*(\Omega) = \sqrt{T_{BS}}\,y(\Omega) + \sqrt{R_{BS}}\,v(\Omega). \tag{41}$$

We have confirmed that the results for the QN curves using Eq. (41) are consistent with those achieved with Eqs. (38) and (39). For the case in the CC-SR scheme, as illustrated in Fig. 2(b), we assume that the distance from the inserted BS to $M_{SR}$ is negligible, and represent the field operators just after the beam splitter as $e^N(\Omega)$ and $f^N(\Omega)$ and the vacuum fields as $p(\Omega)$ and $q(\Omega)$. We have the following relationships:

$$e^N(\Omega) = \mathcal{R}_{AUX}\tilde{e}^N(\Omega), \tilde{f}^N(\Omega) = \mathcal{R}_{AUX}f^N(\Omega), \tag{42a}$$

$$e(\Omega) = \sqrt{T_{BS}}\,e^N(\Omega) + \sqrt{R_{BS}}\,p(\Omega), f^N(\Omega) = \sqrt{T_{BS}}\,f(\Omega) - \sqrt{R_{BS}}\,q(\Omega). \tag{42b}$$

The final input-output relations in Eqs. (25), (26c) and (26d) are then modified as

$$b^N(\Omega) = X^N(\Omega)a^N(\Omega) + Y^N(\Omega)\tilde{h}(\Omega) + P^N(\Omega)p(\Omega) + Q^N(\Omega)q'(\Omega), \tag{43}$$

$$q'(\Omega) = \mathcal{R}_{AUX}q(\Omega), W(\Omega) = \left[\left(\sqrt{T_{BS}}\,\mathcal{R}_{AUX}\right)^{-1} - r_{AUX}\sqrt{T_{BS}}\,\mathcal{R}_{AUX}U(\Omega)\right]^{-1}, \tag{44a}$$

$$X^N(\Omega) = t_{AUX}^2\sqrt{T_{BS}}\,\mathcal{R}_{AUX}U(\Omega)W(\Omega) - r_{AUX}I, \tag{44b}$$

$$Y^N(\Omega) = t_{AUX} r_{AUX} T_{BS} \mathcal{R}_{AUX} U(\Omega) W(\Omega) \mathcal{R}_{AUX} V(\Omega) + t_{AUX} \sqrt{T_{BS}} \mathcal{R}_{AUX} V(\Omega), \quad (44c)$$

$$P^N(\Omega) = t_{AUX} \sqrt{R_{BS}} \mathcal{R}_{AUX} U(\Omega) W(\Omega) \mathcal{R}_{AUX}^{-1}, \quad (44d)$$

$$Q^N(\Omega) = -t_{AUX} r_{AUX} \sqrt{T_{BS} R_{BS}} \mathcal{R}_{AUX} U(\Omega) W(\Omega) - t_{AUX} \sqrt{R_{BS}} I. \quad (44e)$$

Here the superscript "$N$" stands for including the QN from the NDM. The noise spectral density of the CC-SR considering the excess noise from the WLC is then

$$S_h^\xi(\Omega) = \frac{h_{SQL}^2(\Omega)}{|Y_1 \sin(\xi) + Y_2 \cos(\xi)|^2} \begin{bmatrix} |X_{11}^N \sin(\xi) + X_{21}^N \cos(\xi)|^2 + |X_{12}^N \sin(\xi) + X_{22}^N \cos(\xi)|^2 \\ + |P_{11}^N \sin(\xi) + P_{21}^N \cos(\xi)|^2 + |P_{12}^N \sin(\xi) + P_{22}^N \cos(\xi)|^2 \\ + |Q_{11}^N \sin(\xi) + Q_{21}^N \cos(\xi)|^2 + |Q_{12}^N \sin(\xi) + Q_{22}^N \cos(\xi)|^2 \end{bmatrix}. \quad (45)$$

Without the SSA, the numbers $\sqrt{T_{BS}}$ and $\sqrt{R_{BS}}$ in the above equations are replaced by matrices:

$$t_n = \frac{1}{2} \begin{pmatrix} \sqrt{T_{BS}(\omega_0 + \Omega)} + \sqrt{T_{BS}(\omega_0 - \Omega)} & \sqrt{T_{BS}(\omega_0 + \Omega)} - \sqrt{T_{BS}(\omega_0 - \Omega)} \\ \sqrt{T_{BS}(\omega_0 + \Omega)} - \sqrt{T_{BS}(\omega_0 - \Omega)} & \sqrt{T_{BS}(\omega_0 + \Omega)} + \sqrt{T_{BS}(\omega_0 - \Omega)} \end{pmatrix}, \quad (46)$$

$$\mathfrak{r}_n = \frac{1}{2} \begin{pmatrix} \sqrt{R_{BS}(\omega_0 + \Omega)} + \sqrt{R_{BS}(\omega_0 - \Omega)} & \sqrt{R_{BS}(\omega_0 + \Omega)} - \sqrt{R_{BS}(\omega_0 - \Omega)} \\ \sqrt{R_{BS}(\omega_0 + \Omega)} - \sqrt{R_{BS}(\omega_0 - \Omega)} & \sqrt{R_{BS}(\omega_0 + \Omega)} + \sqrt{R_{BS}(\omega_0 - \Omega)} \end{pmatrix}, \quad (47)$$

and the input-output relations are then

$$\boldsymbol{b}^N(\Omega) = X^N(\Omega) \boldsymbol{a}^N(\Omega) + Y^N(\Omega) \tilde{h}(\Omega) + P^N(\Omega) \boldsymbol{p}(\Omega) + Q^N(\Omega) \boldsymbol{q}(\Omega), \quad (48)$$

$$W(\Omega) = \left[ \left( t_n \mathcal{R}_{AUX} \right)^{-1} - r_{AUX} \mathcal{R}_{AUX} t_n U \right]^{-1}, \quad (49a)$$

$$X^N(\Omega) = t_{AUX}^2 \mathcal{R}_{AUX} t_n U(\Omega) W(\Omega) - r_{AUX} I, \quad (49b)$$

$$Y^N(\Omega) = t_{AUX} r_{AUX} t_n \mathcal{R}_{AUX} U(\Omega) W(\Omega) \mathcal{R}_{AUX} t_n V(\Omega) + t_{AUX} \mathcal{R}_{AUX} t_n V(\Omega), \quad (49c)$$

$$P^N(\Omega) = t_{AUX} \mathcal{R}_{AUX} t_n U(\Omega) W(\Omega) \mathcal{R}_{AUX}^{-1} t_n^{-1} \mathfrak{r}_n, \quad (49d)$$

$$Q^N(\Omega) = -t_{AUX} r_{AUX} \mathcal{R}_{AUX} t_n U(\Omega) W(\Omega) \mathcal{R}_{AUX} \mathfrak{r}_n - t_{AUX} \mathcal{R}_{AUX} \mathfrak{r}_n. \quad (49e)$$

## IV. NOISE DENSITY CURVES FOR THE CC-SR CONFIGURATION

Assuming the GW detector is working at $I_0 = I_{SQL}$, we plot $h_n(\Omega)/h_{SQL}(\gamma)$ for two quadratures $b_1$

($\xi = \pi/2$) and $b_2$ ($\xi = 0$). The noise curves for the SR configuration are plotted as red solid (dashed) for $b_1$ ($b_2$) in Fig. 4 [both (a) and (b)], generated by using the approximation $\Omega L/c \ll 1$ and $t_1 \ll 1$ in Eqs. (19a)–(19b), and setting $r_{AUX} = 0$ and $L_{AUX} = 0$ in Eqs. (26c)–(26d), in agreement with the results in Ref. 7. The corresponding noise curves for the CC-SR configuration of Fig. 2, but *without the NDM*, are shown by green solid (dashed) lines for the $b_1$ ($b_2$) quadrature, for two different values of $R_{AUX} (\equiv r_{AUX}^2)$: 99% in Fig. 4(a) and 99.5% in Fig. 4(b). Both curves have a minimum at around $\Omega_c/(2\pi) = 127$Hz; this value is determined by the choice of $L_{AUX}$. As can be seen, use of a higher reflectivity $M_{AUX}$ reduces the QN at the minimum (i.e. increases sensitivity) but narrows the width of the resonance dip.

Next, we show the noise curves for the CC-SR configuration of Fig. 2, in the presence of the NDM and under the SSA, but without taking into account the QN from the NDM. The $b_1$ ($b_2$) quadrature is shown as the black solid (dashed) lines in Fig. 4, again for two different values of $R_{AUX}$. As expected, the addition of the NDM causes the WLC effect, thus broadening the dips significantly, covering the range from ~100Hz to ~5000Hz without considering the excess QN from the NDM, for $R_{AUX} = 99\%$, without reducing sensitivity. This is consistent with the result for the classical response of the CC-SR presented in Sec. II B. For the higher value of $R_{AUX} = 99.5\%$, we also see a significant broadening (~200Hz to ~4000Hz), again without a reduction in sensitivity. When the excess QN from the NDM is taken into account, the noise curves get narrower, as shown in purple. When the SSA is relaxed, the results are shown in black in Figs. 5(a) and 5(b) for the two values of $R_{AUX}$ above. We find that the noise curves for $b_1$ change a lot, while those for $b_2$ are modified slightly with a narrower width and higher sensitivity. The noise curve for $b_2$ when $R_{AUX} = 99\%$ is highly broadened with the introduction of the WLC [plotted as the purple dashed line in Fig. 5(a)]. Even though the curves remain above the SQL, this broadening is a very important result [21] and may prove useful in the aLIGO, since the noise floor in the current design is above the SQL anyway [7]. The fact that the noise floor remains above the SQL when the QN from the NDM is taken into account is not a fundamental constraint. As we will show soon, when the dispersion profile of the NDM is tailored to take into account the effect of OM resonance, it is possible to get the sensitivity well below the SQL.

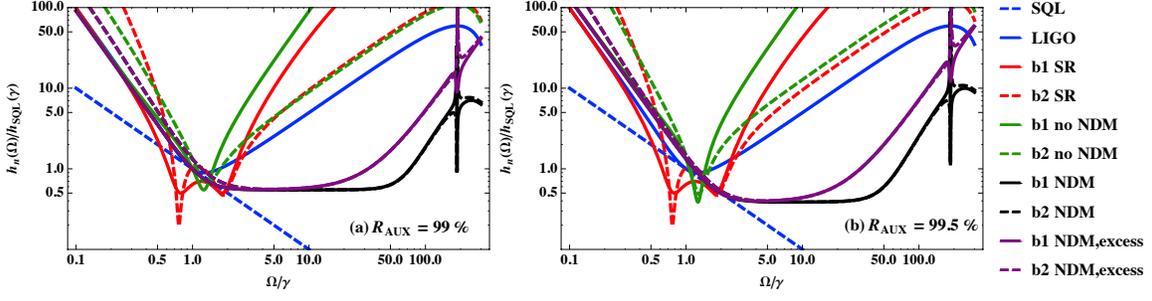

FIG. 4. Log-log plots of $h_n(\Omega)/h_{SQL}(\gamma)$ versus $\Omega/\gamma$ for the two quadratures $b_1$ and $b_2$, for the CC-SR configuration when (a) $R_{AUX} = 99\%$ and (b) $R_{AUX} = 99.5\%$, with (without) the NDM are shown in black (green) under the SSA. The noise curves considering the QN from the NDM are shown in purple. The noise curves for $b_{1,2}$ in the SR configuration with $R_{SR} = 81\%$, $\varphi_{SRC} = \pi/2 - 0.47$ and $\Phi_{SRC} = 0$, are shown as the red curves. The noise curve for LIGO and the SQL line are plotted in blue.

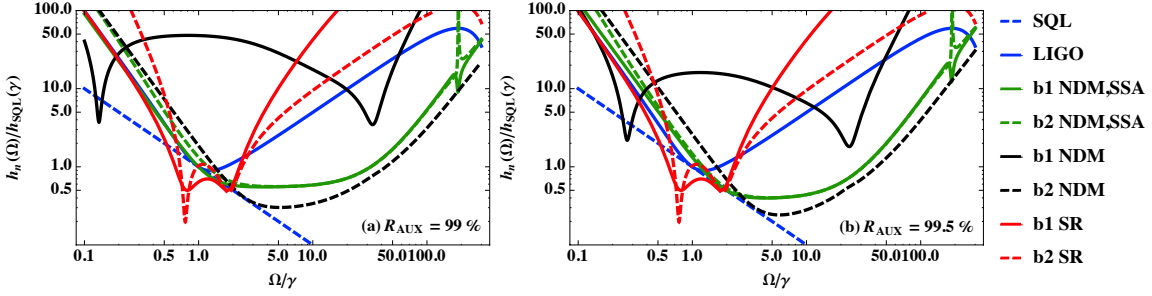

FIG. 5. Log-log plots of $h_n(\Omega)/h_{SQL}(\gamma)$ versus $\Omega/\gamma$ for the two quadratures $b_1$ and $b_2$, for the CC-SR configuration when (a) $R_{AUX} = 99\%$ and (b) $R_{AUX} = 99.5\%$, with and without the SSA.

In the SR configuration, the OM resonance dips are induced by the phase $\vartheta_{SRC}$ gained from reflection off the SRC. It is shown in Ref. 7 that the position of the dips in the noise curves with a high $r_{SR}$ agrees well with the resonances of the closed system ($r_{SR} = 1$) with no GW signal [$h(\Omega) = 0$]. We follow the method in Ref. 7 to evaluate the free oscillation modes for the closed system. Similarly to the quantum-field operators for the two quadrature fields, we consider a classical field $E$ consisting of two quadrature components $E_1$ and $E_2$, i.e. $\boldsymbol{E}(\Omega) = (E_1(\Omega), E_2(\Omega))^T$. $\boldsymbol{E}$ enters Port B and returns as $\boldsymbol{E}'$ after propagating through the two arms. At resonance, $\boldsymbol{E}'$ propagates round-trip in the SRC and returns in phase with $\boldsymbol{E}$. As a result we find

$$\left[ A(\Omega) - \mathcal{R}_{SRC}^{-2} \right] E(\Omega) = \mathbf{0}, \tag{50}$$

where $A(\Omega)$ is as defined in Eqs. (18) and (19a), and $\mathcal{R}_{SRC}$ is as defined in the paragraph preceding Eq. (24a). Therefore, the characteristic equation for this system is

$$\left| A(\Omega) - \mathcal{R}_{SRC}^{-2} \right| = 0. \tag{51}$$

The solution of this equation yields the eigenvalue $\varphi_{SRC}$ (denoted by $\varphi_{SRC}^0$) in the limit $\Omega L_{SRC}/c \ll \pi$, so that $\Phi_{SRC} \approx 0$. As is discussed in Sec. II, only one of the two sidebands will be on resonance for a specific choice of $\varphi_{SRC}$, and whether the plus- or minus- sideband is on resonance depends on the value of $\varphi_{SRC}$. Therefore, the phase shift experienced by a beam inside the arm cavities upon reflection from the SRC can be expressed as

$$\vartheta_{SRC} = Arg\left( \frac{t_1^2 r_{SR} e^{2i\varphi_{SRC}^0}}{1 + r_1 r_{SR} e^{2i\varphi_{SRC}^0}} + r_1 \right). \tag{52}$$

While we derived Eq. (52) for the limit $r_{SR} = 1$, the equation is still valid for a large value of $r_{SR}$, as discussed in Ref. 7.

In the CC-SR the SRC is tuned to resonance of the carrier wavelength and effectively disappears for the range of GW sidebands of interest. Therefore, the frequency-dependent phase $\vartheta_{SRC}$ can be effectively achieved by round-trip propagation in the cavity of length $L_{AUX}$ with a dispersive medium, i.e. $2(1+\chi'/2)kL_{AUX}(\mod 2\pi) = \vartheta_{SRC}$. The dispersion is centered around the sideband in OM resonance in the same system but without the dispersive medium, whose frequency is determined by $L_{AUX}$. However, the exact dispersion required by Eq. (52) is hard to achieve, and we first use the Lorentzian model described by Eqs. (4) and (5) as an approximation to Eq. (52) for a certain range of frequencies.

The results for the QN-limited sensitivity using an NDM are plotted in Fig. 6, where the OM resonance is located at $\omega_c = \omega_0 - \Omega_c$ [$\Omega_c/(2\pi) = 236.3\text{Hz}$] without the NDM for a specifically chosen $L_{AUX}$, and the dispersion is centered at $\omega_c$. When $R_{AUX} = 99.9\%$, the QN curves (shown as black curves) show a sub-SQL region around $\omega_c$. Here the parameters for the NDM are chosen so that at the center of the dispersion $\omega_c$, the gain is exactly zero, that is $\chi''(\omega_c) = 0$ and $g(\omega_c) = 1$ [here $\Gamma_e/(2\pi) \approx 400\,\text{Hz}$]. The minimum of the noise is $\sim 0.18 h_{SQL}(\gamma)$, and the bandwidth of the sub-SQL region is ~50Hz. Compared with the curve in the SR configuration with the highest sensitivity, which occurs for the second quadrature $b_2$ (plotted as a dashed red line), the minimum is comparable while the bandwidth is much larger, resulting in an improvement in sensitivity-bandwidth product by a factor of ~7. We also show the results with the SSA in green as a comparison. We see that the SSA causes only a slight modification in the QN curves in this case.

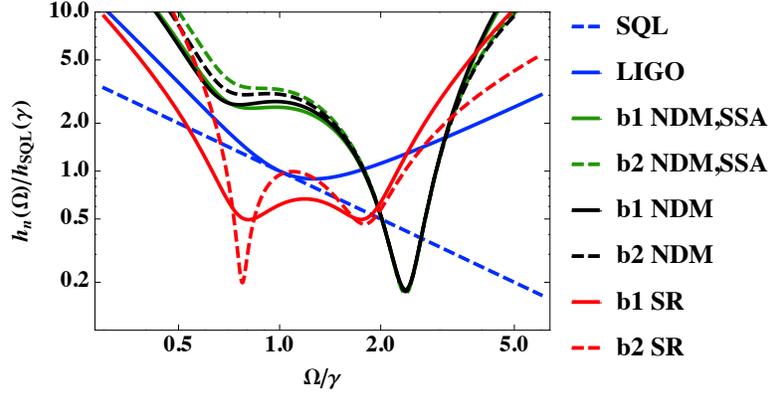

FIG. 6. Log-log plots of $h_n(\Omega)/h_{SQL}(\gamma)$ versus $\Omega/\gamma$ for the CC-SR scheme with a modified dispersion of the NDM centered at $\omega_c = \omega_0 - \Omega_c$ [$\Omega_c/(2\pi) = 236.3\,\text{Hz}$]. The plots with (green) and without (black) the SSA are shown for comparison.

In order to achieve a broader sub-SQL region, we center the dispersion at $\omega_c = \omega_0 - \Omega_c$ [$\Omega_c/(2\pi) = 200\,\text{Hz}$] and tailor it according to Eq. (52). $L_{AUX}$ is also modified so that the OM resonance without the NDM is moved to the new $\omega_c$. The QN curves with $R_{AUX} = 99.9\%$ under the SSA are shown in green in Fig. 7 [here $\Gamma_e/(2\pi) \approx 16\,\text{kHz}$], which exhibits a sub-SQL region of ~100Hz in width around $\omega_c$, with the minimum beating the SQL by a factor of 5. Without the SSA, the curves are plotted in black, exhibiting a sub-SQL region somewhat narrower than the case with SSA. The QN curves for both quadratures show resonance dips around $\Omega_c$, and the dip for $b_1$ is broader but shallower than that for $b_2$. If we compare the dip for $b_1$ in this case with the highest sensitivity dip (corresponding to $b_2$) in the SR case, we see that the sensitivity-bandwidth product is enhanced by a factor of ~14.

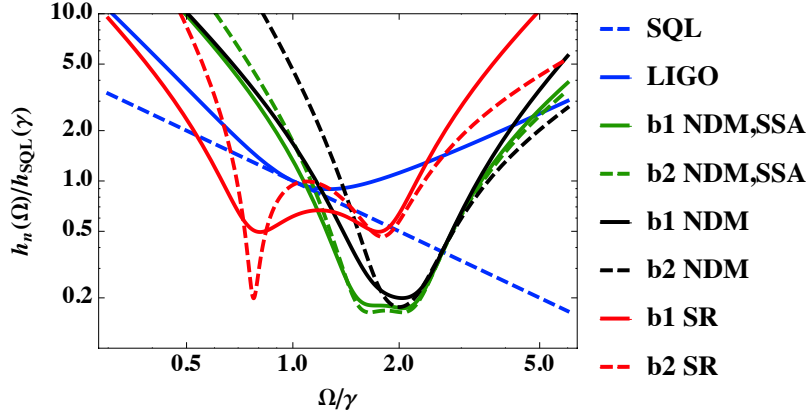

FIG. 7. Log-log plots of $h_n(\Omega)/h_{SQL}(\gamma)$ versus $\Omega/\gamma$ for the CC-SR scheme with a modified dispersion of the NDM centered at

$$\omega_c = \omega_0 - \Omega_c \ [\ \Omega_c/(2\pi) = 200 \text{Hz} \ ].$$

## V. WLC-SR CONFIGURATION

In the preceding section, we have shown that it is indeed possible to broaden the QN-limited response without a reduction in sensitivity. However, the degree of broadening is significantly smaller than the same found in the classical response. To overcome this limitation of broadening, we consider next an alternative scheme, simpler than the CC-SR, where we insert a dispersive medium in the SRC in aLIGO [Fig. 8(a)], with the propagation phase in the SRC approximating the eigenvalues $\varphi_{SRC}^0$ determined by Eq. (51) . We set $L_{SRC}$ to ~10m and assume that the dispersive medium fills up the whole SR cavity. The QN can be calculated by taking $r_{AUX} = 0$ and $L_{AUX} = 0$, and $\mathcal{R}_n(\varphi_{SRC}, \Phi_{SRC}) = \mathcal{R}_{SRC}$ in Eqs. (25), (26a)−(26d), while the QN including the QN from the dispersive medium can be calculated using the method in Sec. III C [see Fig. 8(b)].

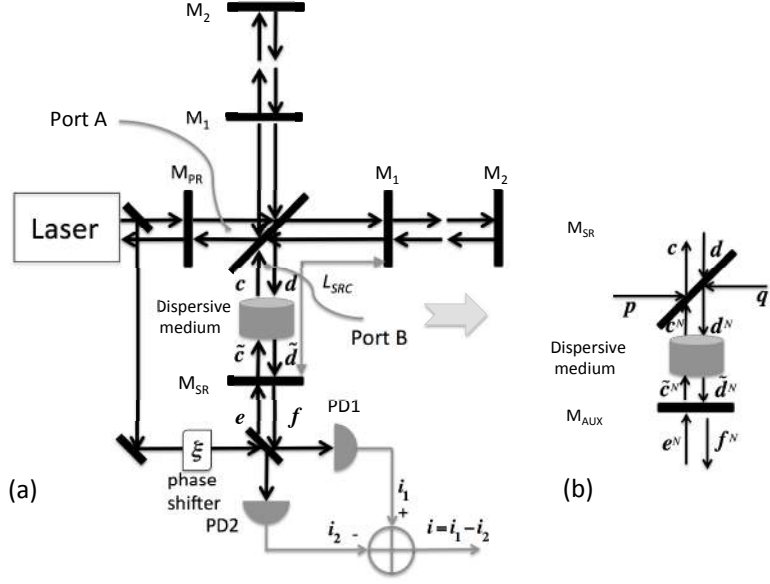

FIG. 8. (a) WLC-SR design. A dispersive medium is inserted in the SRC to achieve a broader sub-SQL dip. (b) Schematic view of the SR cavity with excess QN modeled by inserting a beam splitter with power reflectivity $R_{BS}$ and power transmissivity $T_{BS}$. Here $p$ and $q$ are the vacuum noises that leak into the system.

### A. Phase compensation using a positive dispersion medium

As a direct comparison to the SR scheme in Ref. 7, where there is an OM resonance at $\omega_c = \omega_0 - \Omega_c$ [$\Omega_c/(2\pi) = 77.5\text{Hz}$], we first choose $L_{SRC}$ so that the OM resonance condition is satisfied at $\omega_c$ and center the dispersion there. A careful inspection of the frequency dependence of the phase $\varphi_{SRC}^0$ shows that in order to compensate for it, one must make use of positive dispersion. Of course, the concept of using a WLC to broaden the response of a cavity has traditionally been based on the use of negative dispersion, due to the nature of the round-trip phase in a conventional cavity. However, as we see here, for a more complex system, this general notion does not necessarily hold. For the PDM we need to use here, we model the dispersion as a narrow transparency peak on top of a broader absorption dip, opposite to the NDM described in Eqs. (4) and (5). In the limit of vanishing Rabi frequencies ($\Omega_k \to 0$), $\chi'$ and $\chi''$ of the PDM necessary to achieve the phase $\varphi_{SRC}^0$ are plotted in Fig. 9, with the parameters chosen such that at the center of the dispersion $\chi''(\omega_c) = 0$ and $g(\omega_c) = 1$. The PDM can be realized, for example, via EIT. It is shown in Ref. 16 that the QN of the Λ-type EIT system can be correctly described by the single-channel Caves model. The QN-limited sensitivity with the QN from the PDM is shown in Fig. 10. When the power reflectivity of $M_{SR}$ is $R_{SR} = 81\%$ (same as the SR case), the sensitivity curve for $b_2$ under the SSA [plotted as dashed green lines in Fig. 10(a)] exhibits a sub-SQL region 3 times broader than that for $b_2$ in the case of an empty SRC (plotted as a red dashed line) without decrease in sensitivity. When $R_{SR} = 97\%$ [Fig. 10(b)], the sensitivity increases. The minimum value for the second quadrature $b_2$ decreases by a factor of 2 with a loss in the bandwidth, and the curve for the first quadrature $b_1$ is lowered to about the same level as that for $b_2$ in the SR case. Without the SSA, the QN curves are shown in black. In this case, however, the QN is lifted up to $\sim 0.5 h_{SQL}(\gamma)$ for both $R_{SR} = 81\%$ and $97\%$, and there is no improvement in the sensitivity-bandwidth product.

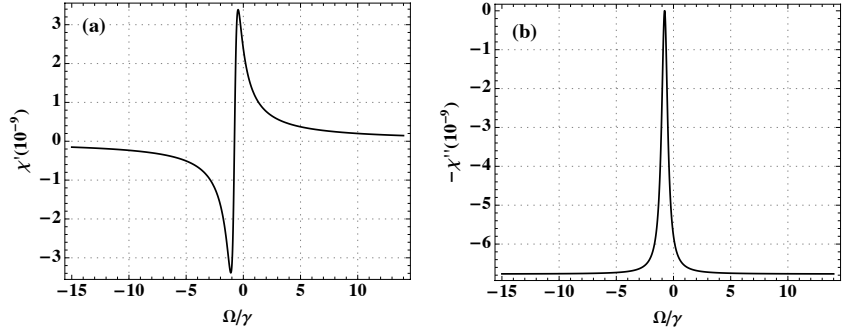

FIG. 9. Plots of (a) $\chi'$ and (b) $-\chi''$ versus $\Omega/\gamma$ for the PDM used in plotting Fig. 10.

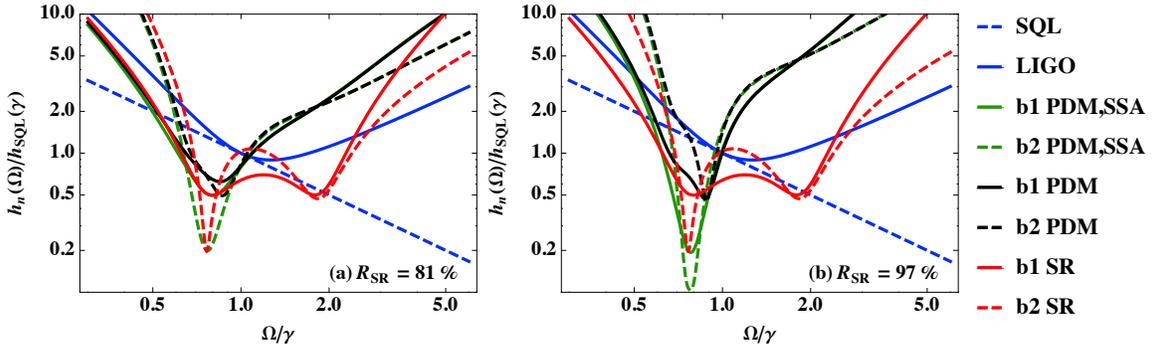

FIG. 10. Log-log plot of $h_n(\Omega)/h_{SQL}(\gamma)$ versus $\Omega/\gamma$ in the WLC-SR configuration with a tailored positive dispersion centered at $\omega_c = \omega_0 - \Omega_c$ [ $\Omega_c/(2\pi) = 77.5 \text{Hz}$ ] when (a) $R_{SR} = 81\%$ and (b) $R_{SR} = 97\%$.

## B. Phase compensation using a negative dispersion medium

We next consider a case where the dispersion is centered at a higher frequency. For this case, an NDM has to be used, whose $\chi'$ and $\chi''$ are plotted in Fig. 11 for $\Omega_c/(2\pi) = 200$Hz [here $\Gamma_e/(2\pi) \approx 16$kHz]. The results for the QN limited sensitivity curves are shown in Fig. 12. Here $L_{SRC}$ is changed so that the OM resonance of the system without the dispersive medium is at $\omega_c = \omega_0 - \Omega_c$. We choose the parameters for the dispersion such that $\chi''(\omega_c) = 0$ and $g(\omega_c) = 1$. Under the SSA, the noise curves exhibit a rather broad sub-SQL region of a bandwidth around 140Hz with its minimum ~5.5 times smaller than $h_{SQL}(\gamma)$ when $R_{SR} = 97\%$. We also show in Fig. 12 that when the SSA is removed, the results remain almost unchanged with the valley ~5Hz narrower in width. To summarize, the sensitivity-bandwidth product is enhanced by nearly a factor of 18 compared to the highest sensitivity result (for $b_2$) in the SR scheme.

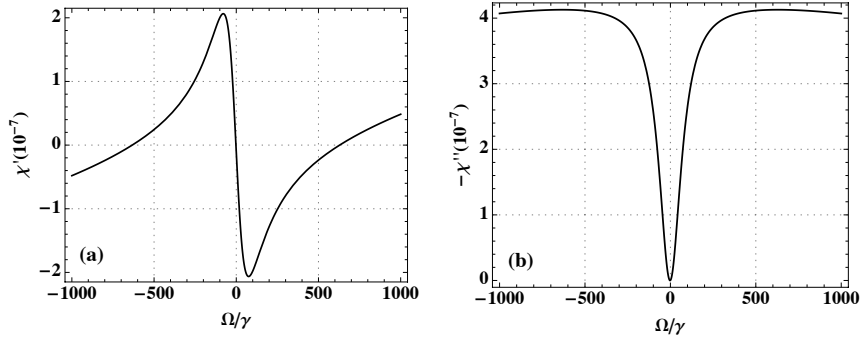

FIG. 11. Plots of (a) $\chi'$ and (b) $-\chi''$ versus $\Omega/\gamma$ for the NDM used in plotting Fig. 12.

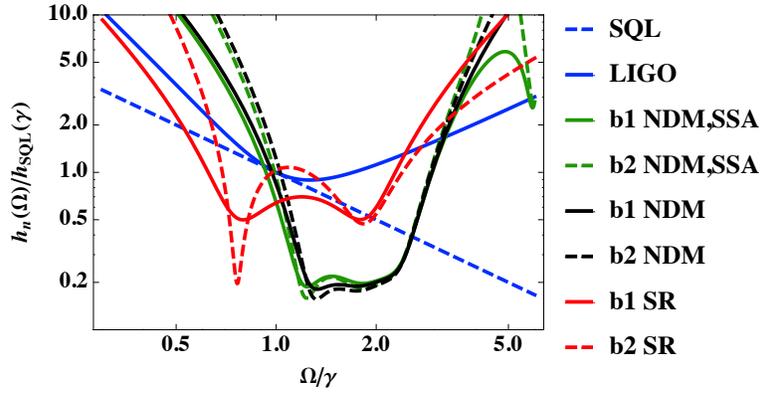

FIG. 12. Log-log plot of $h_n(\Omega)/h_{SQL}(\gamma)$ versus $\Omega/\gamma$ in the WLC-SR configuration with a tailored negative dispersion centered around $\omega_0 - \Omega_c$ ($\Omega_c/(2\pi) = 200$Hz). The noise curves when $R_{SR} = 97\%$ without (with) the SSA are plotted in black (green).

### C. Lasing condition

When the gain medium is introduced, one potential issue is that the system might start lasing. Consider the cavity composed by M$_{SR}$ and the arm cavity as a compound mirror M$_{12}$, which entails an effective quality factor $Q_c = 3.8 \times 10^{13}$. Since the OM effects modify the resonance position to $\omega_0 - \Omega_c$ [$\Omega_c/(2\pi) = 200$Hz] in the system in the case shown above in Sec. V B, we alter the length $L_{SRC}$ so that the semi-classical resonance of the cavity is located at $\omega_0 - \Omega_c$. In steady state, the phase and amplitude of the field inside the cavity satisfy a set of self-consistent equations [19]:

$$\left(1 + \frac{1}{2}\chi'(E,\omega)\right)\frac{\omega}{c}2L_{SRC} + \vartheta_{12}(\omega) = \frac{\omega_{res}}{c}2L_{SRC} + \vartheta_{12}(\omega_{res}), \tag{53}$$

$$\chi''(E,\omega) = -1/Q_c. \tag{54}$$

where $\vartheta_{12}$ is the frequency-dependent phase that the field gains from reflecting off M$_{12}$, $\omega$ is the lasing frequency, and $\omega_{res} = 2\pi f_{res}$ is the resonant frequency of the cavity in the absence of the medium. For the PDM with $\chi''(\omega_c) = 0$, $\chi''$ is always positive for all frequencies, therefore the system is always below the lasing threshold. For the NDM, the boundaries of the lasing range can be solved by setting $E = 0$ ($\Omega_k = 0$) in Eq. (4) and plugging the resulting value of $\chi''$ into Eq. (54). For frequencies $f_{-1} < f < f_1$, $f < f_{-2}$ or $f_2 < \omega$ ($f_{\pm 1} = f_{res} \pm 2.0$Hz, $f_{\pm 2} = f_{res} \pm 2.0 \times 10^9$Hz, as shown in Fig. 13), the gain cannot compensate for the cavity loss, so that $E = 0$; otherwise Eq. (54) is satisfied, from which we can solve for $E(f)$ as a function of frequency $f = \omega/(2\pi)$. It can be seen that the frequency $\omega$ that satisfies Eq. (53) falls within the range $f_{-1} < f < f_1$, where the gain is below lasing threshold and $E = 0$. Thus, we find that lasing will not occur in this system.

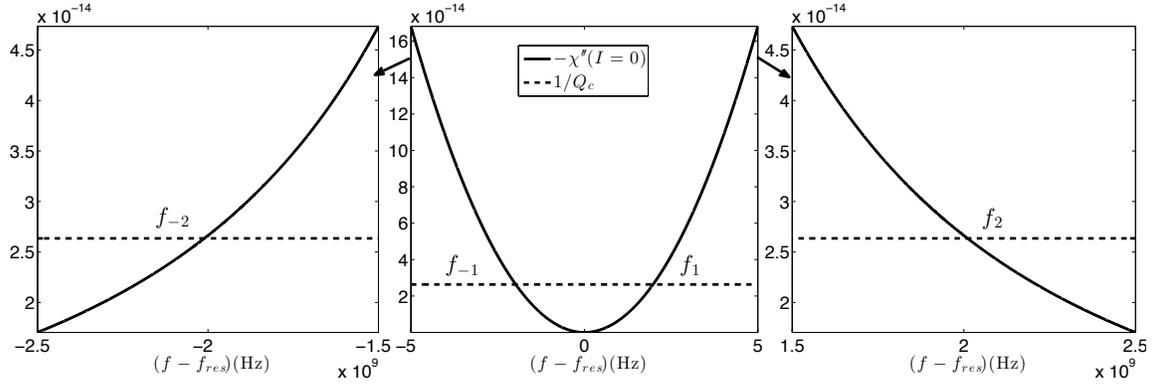

FIG. 13. Illustration of the lasing ranges.

## VI. CONSIDERATION OF AN EXPLICIT SYSTEM FOR THE NDM AND MORE EXACT CONSIDERATION OF THE QN

As we have discussed briefly earlier in the paper, the Caves model makes some assumptions that may not necessarily hold for some systems. In Ref. 16, we have carried out a comprehensive and systematic analysis in order to determine whether the noise in a particular system can be predicted correctly by using the Caves model. Specifically, we have used a Master Equation (ME) approach to determine the noise spectrum for a range of excitations, involving two or more energy levels, under conditions that may yield gain or attenuation. For each case, we have then computed the noise spectrum using the Caves model. In some cases, we have found these models to agree with each other. In other cases, we have shown that the details of the process must be considered to compute the noise using the ME while the Caves model cannot be used. This is true, for example, in a composite system where an inverted two-level transition produces gain, while a non-inverted two-level system produces absorption, with the gain exactly canceling the absorption at a particular probe frequency. While a naive, single-channel Caves model would imply no noise at this frequency, the ME result predicts a noise that is substantial at this frequency. We have also shown that in a $\Lambda$-type EIT (electromagnetically induced transparency) system where, in the steady state, the atoms are in a so-called dark-state, representing a superposition of metastable ground states, and no population in excited states, the single-channel Caves model yields the correct result. Such a system occurs, for example, in a $\Lambda$-type EIT system. Inspired by EIT, we have shown that it is also possible to produce such an EIT system where the steady state is essentially a dark state, in a five-level transition which produces a broad gain away from the EIT condition. This configuration, which we call a Gain-EIT (GEIT) system, can be tailored to produce the negative dispersion necessary for realizing the WLC effect. Here, we first describe this GEIT system briefly before considering its application as the NDM in the WLC-SR configuration. More details about this system can be found in Ref. 16.

The GEIT system is shown schematically in Fig. 14. It is a five-level M-system, where the transitions $|1\rangle$-$|4\rangle$, $|2\rangle$-$|4\rangle$ and $|3\rangle$-$|5\rangle$ are coupled by the pump fields $\Omega_1$, $\Omega_2$ and $\Omega_4$, respectively, while the transition $|2\rangle$-$|5\rangle$ is coupled by the probe field $\Omega_3$. We assume that $\delta_i (i=1,2)$ is chosen to balance the differential light shift experienced by levels $|1\rangle$ ($\Omega_1^2/(4\delta_1)$) and $|2\rangle$ ($\Omega_2^2/(4\delta_2) + \Omega_3^2/(4\delta_3)$), so that the left leg of the M-system composed by $|1\rangle - |4\rangle - |2\rangle$ is resonant. For the other leg, $|2\rangle - |5\rangle - |3\rangle$, we define $\delta_3 = \delta_{30} + \Delta$, where $\Delta = 0$ corresponds to the condition where the differential light shift experienced by levels $|3\rangle$ ($\Omega_4^2/(4\delta_4)$) and $|2\rangle$ ($\Omega_2^2/(4\delta_2) + \Omega_3^2/(4\delta_3)$) is balanced.

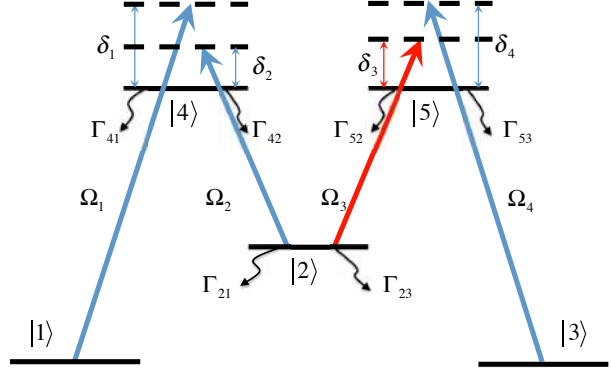

FIG. 14.   Schematic illustration of the five-level GEIT system.

We consider the case when $\gamma/(2\pi) = 6\text{MHz}$, $\Omega_1 = \gamma$, $\Omega_2 = 10^2\gamma$, $\Omega_3 = 10^{-6}\gamma$, $\Omega_4 = 10^{-1}\gamma$, and $\delta_1 \approx \delta_2 \approx \delta_3 \approx \delta_4 \approx 10^3\gamma$. We show in Fig. 15(a) that a transmission profile with a dip on top of a broad gain peak is produced and the negative dispersion is plotted in Fig. 15(b). We have also verified that the gain remains linear (i.e. independent of the amplitude of $\Omega_3$) as the amplitude of $\Omega_3$ approaches a vanishing value.

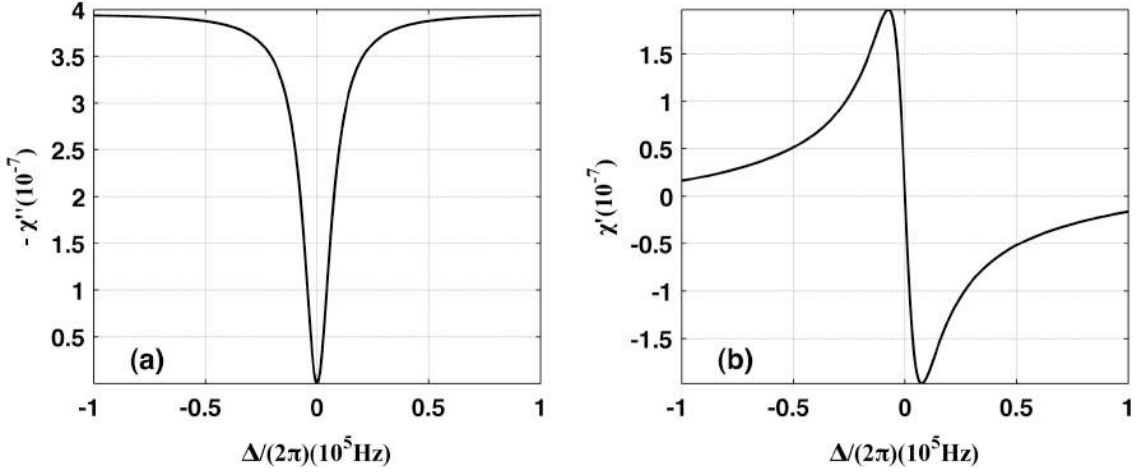

FIG. 15. Plots of (a) $-\chi''$ and (b) $\chi'$ versus $\Delta/2\pi$ for the GEIT system in Fig. 14. Here, $\gamma/(2\pi) = 6$MHz, $\Gamma_{41} = \Gamma_{42} = \Gamma_{52} = \Gamma_{53} = \Gamma_{21} = \Gamma_{23} = \gamma/2$, $\delta_1 \approx \delta_2 \approx \delta_3 \approx \delta_4 \approx 10^3\gamma$, $\Omega_1 = \gamma$, $\Omega_2 = 10^2\gamma$, $\Omega_3 = 10^{-6}\gamma$ and $\Omega_4 = 10^{-1}\gamma$.

To evaluate the QN using the result from the ME model [16], we need to make use of the equations

$$y^*(\Omega) = \sqrt{g}y(\Omega) + X_1 v_1^\dagger(\Omega) + X_2 v_2(\Omega), \tag{55}$$

$$X_1 = \sqrt{(g-1)\frac{G_1}{G_1 - G_2}}, X_2 = \sqrt{(g-1)\frac{G_2}{G_1 - G_2}}, \tag{56}$$

instead of Eqs. (40) and (41), where $v_1^\dagger$ and $v_2$ are vacuum fields that account for the additional noise. Here, $G_1$ and $G_2$ are the contributions of the amplification and attenuation, respectively, to the net gain $g = \exp(G_1 - G_2)$, which are proportional to $\mathcal{A}$ and $\mathcal{B}$ in Ref. 16. The values of $\mathcal{A}$ and $\mathcal{B}$ are calculated from solving the master equation of the GEIT system.

Using the results from the ME model and Eqs. (55) and (56), we can calculate the QN-limited sensitivity. We show in Fig. 16 that the sensitivity curves using the GEIT with the parameters same as in Fig. 15 are very similar to the noise curves we plot in Fig. 12. The curves remain well below the SQL, and they have an enhancement of sensitivity-bandwidth product by a factor of 16.55 compared to the curve in the SR configuration with the highest sensitivity. As a comparison, we show in Fig. 17 the sensitivity curves for the first quadrature, when the QN from the NDM is taken into account using the ME approach and the Caves model, respectively. In this case, the results predicted by these models differ by less than 0.2%, and the difference is not noticeable. Similar agreement is seen for the second quadrature as well (not shown).

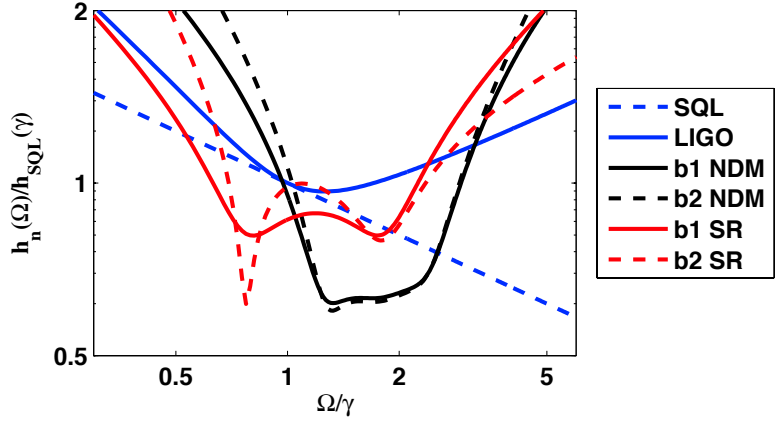

FIG. 16.    Log-log plot of  $h_n(\Omega)/h_{SQL}(\gamma)$  versus  $\Omega/\gamma$  in the WLC-SR using the GEIT system with the parameters in Fig. 15 as the NDM.

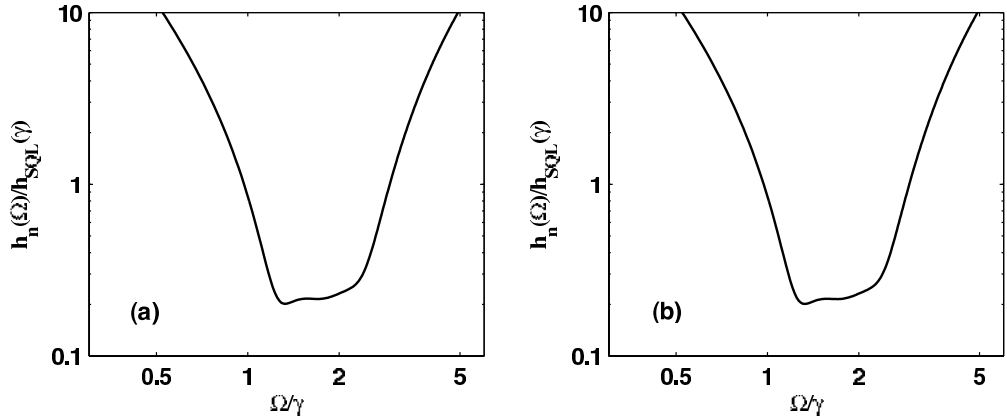

FIG. 17.    Log-log plot of  $h_n(\Omega)/h_{SQL}(\gamma)$  versus  $\Omega/\gamma$  in the WLC-SR using the GEIT system with the parameters in Fig. 15 as the NDM. Then QN from the NDM is taken into account using (a) the ME approach and (b) the Caves model.

Using a different set of parameters for the GEIT system with the susceptibilities plotted in Fig. 18, we are able to achieve an even higher enhancement, 17.66, in the sensitivity-bandwidth product. The QN-limited sensitivity curves are shown in Fig. 19. In this case, the sensitivity predicted by the Caves model differs significantly from the result determined by the ME approach. At the bottom of the sensitivity curves, the difference is about 13%. Therefore, in general, the QN must be calculated by the ME approach only. Finally, it can be shown that lasing will not occur in this GEIT system by carrying out an analysis similar to what is described in Sec. V C. Thus, the predicted enhancement in sensitivity, as shown in Fig. 19, is not invalidated by any potential instability due to lasing.

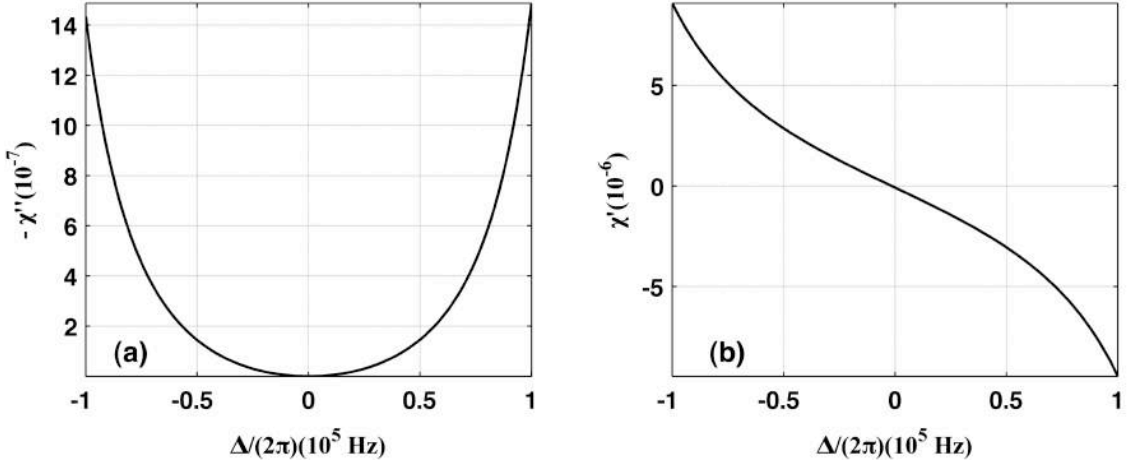

FIG. 18. Plots of (a) $-\chi''$ and (b) $\chi'$ versus $\Delta/2\pi$ for the GEIT system in Fig. 14. Here, $\gamma/(2\pi) = 6\text{MHz}$, $\Gamma_{41} = \Gamma_{42} = \Gamma_{52} = \Gamma_{53} = \gamma/2$, $\Gamma_{21} = \Gamma_{23} \approx 2.02\times 10^{-3}\gamma$, $\delta_1 \approx \delta_2 \approx \delta_3 \approx \delta_4 \approx 10^3\gamma$, $\Omega_1 = \gamma$, $\Omega_2 = 10^2\gamma$, $\Omega_3 = 10^{-6}\gamma$ and $\Omega_4 = 10^{-1}\gamma$.

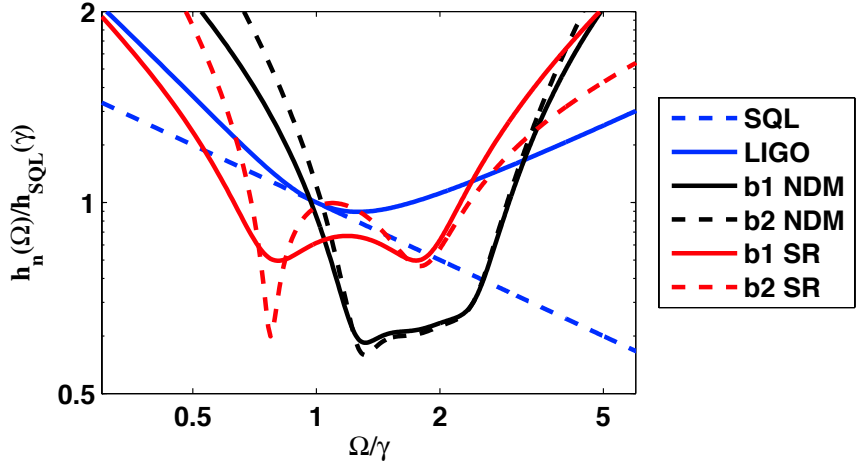

FIG. 19. Log-log plot of $h_n(\Omega)/h_{SQL}(\gamma)$ versus $\Omega/\gamma$ in the WLC-SR using the GEIT system with the parameters in Fig. 18 as the NDM.

It should be possible to demonstrate the five-level GEIT system using sublevels in alkali atoms such as Rb. However, current LIGO operates at the wavelength of 1064nm. We have not yet been able to identify a set of atomic transitions that can be used to realize the GEIT system at this wavelength. It is certainly possible that the operating wavelength of the next-generation LIGO would be chosen to coincide with an alkali atom transition, thus making it possible to implement rather easily the WLC-SR configuration using the GEIT system. For the current wavelength of 1064 nm, one possible scheme for realizing the GEIT involves making use of a set of coupled fiber resonators, along with amplification induced, for example, via stimulated Brillouin scattering [22]. Conventional EIT has already been demonstrated in coupled fiber resonators [23]. Work is currently in progress to devise the GEIT scheme using this approach, and analyze the QN properties of such a system.

## VII. CONCLUSION

We have derived the QN density curves that show minimum detectable GW amplitudes for the CC-SR scheme (Fig. 2) and the WLC-SR scheme (Fig. 8), following the two-photon formalism [13] We first take the QN from the NDM into account using the single-channel Caves model to determine an upper bound of the degree of enhancement in the sensitivity-bandwidth product. In the CC-SR GW detector, the conventional SR mirror is replaced by a CC containing an NDM. We have carried out a detailed QN analysis for various choices of parameters, taking into account all possible sources of QN, including the QN due to the NDM, under the assumption that all excess noise is suppressed below the QN. In keeping with our previous proposal, we first considered the case where the negative dispersion is centered at the semiclassical resonance frequency at which the maximum sideband amplitude is generated in the absence of the NDM, without taking into account the OM effects. In this case, even if the QN from the NDM is taken into account, the QN-limited sensitivity curves exhibit a significant broadening. Although the curves remain above the SQL, this result is of considerable significance, since the current noise in the aLIGO design does not allow operating in the sub-SQL region. We then modify the spectral profile of the dispersion so that it is centered at a different, optimally chosen, frequency $\omega_0 - \Omega_c$ [$\Omega_c/(2\pi) = 200$Hz], which is the position of the OM resonance for the chosen $L_{AUX}$, and the shape of the dispersion curve is tailored to compensate for the nonlinear phase variation induced by the OM effects. Under these conditions, the noise curves fall significantly below the SQL with its minimum beating $h_{SQL}(\gamma)$ by a factor of 5, while retaining a broad bandwidth ~120Hz. This represents an upper bound of ~14 for the factor by which the sensitivity-bandwidth product is increased, compared with the highest sensitivity quadrature ($b_2$) in the SR design [7]. We also considered an alternative, simpler WLC-SR design, which adds an NDM or a PDM in front of the conventional SR mirror in the SR configuration, depending on where the dispersion is centered. The nearly Lorentzian dispersion is tailored to compensate, as closely as possible, the nonlinear phase variation produced by the OM resonance. At the center of dispersion, which is the OM resonance frequency for the chosen SRC length $L_{SRC}$, shifted from the semiclassical resonance frequency mentioned above, the QN due to the dispersive medium is minimal but increases away from this point. After optimization of the various parameters, we have identified conditions using an NDM with $\Omega_c/(2\pi) = 200\text{Hz}$, under which the noise curves beat the SQL by a factor of 5.5. This represents an upper bound of ~18 for the factor by which the sensitivity-bandwidth product is increased, compared with the highest sensitivity quadrature ($b_2$) in the SR design [7]. Finally, we consider an explicit system for realizing the NDM, which is a five-level, M-configuration GEIT system. For this system, we use a rigorous approach, based on Master Equations [16] to calculate the QN from the NDM, so that the resulting prediction about the enhancement in the sensitivity-bandwidth product is definitive, and not simply an upper bound. Using the GEIT system as the NDM in the WLC-SR, we can get an enhancement of the sensitivity-bandwidth product by a factor of 17.66. Further investigation will focus on identifying practical schemes for implementing this concept.

## Acknowledgments

This work was supported by DARPA through the Slow Light program under Grant No. FA9550-07-C-0030, and by AFOSR under Grant No. FA9550-10-1-0228 and No. FA9550-09-1-0652.

**APPENDIX: ABBREVIATIONS**

| Abb | Description | Abb | Description |
|---|---|---|---|
| aLIGO | Advanced LIGO | OM | Optomechanical |
| AUX | Auxiliary | PD | Photodetector |
| CC | Compound cavity | PDM | Positive dispersion medium |
| EIT | Electromagnetically induced transparency | PR | Power recycling |
|  |  | PRC | Power recycling cavity |
| GEIT | Gain-EIT | QN | Quantum noise |
| GW | Gravitational wave | SQL | Standard quantum limit |
| LIGO | Laser Interferometer Gravitational Wave Observatory | SR | Signal recycling |
|  |  | SRC | Signal recycling cavity |
| LO | Local oscillator | SSA | Single sideband approximation |
| ME | Master Equation | WLC | White light cavity |
| NDM | Negative dispersion medium |  |  |